\documentclass[letterpaper,english,reprint,aps]{revtex4-1}
\pdfoutput=1
\usepackage[T1]{fontenc}
\usepackage[latin9]{inputenc}
\usepackage{bm}
\usepackage{amsmath}
\usepackage{amssymb}
\usepackage{graphicx}

\makeatletter

\pdfpageheight\paperheight
\pdfpagewidth\paperwidth

\usepackage{babel}

\makeatother

\usepackage{babel}
\begin{document}
\title{Topological Photonic Tamm-States and the Su-Schrieffer-Heeger Model }
\author{J. C. G. Henriques, T. G. Rappoport, and Y. V. Bludov}
\affiliation{Department and Centre of Physics, and QuantaLab, University of Minho,
Campus of Gualtar, 4710-057, Braga, Portugal}
\author{M. I. Vasilevskiy and N. M. R. Peres}
\affiliation{Department and Centre of Physics, and QuantaLab, University of Minho,
Campus of Gualtar, 4710-057, Braga, Portugal}
\affiliation{International Iberian Nanotechnology Laboratory (INL), Av. Mestre
José Veiga, 4715-330, Braga, Portugal}
\date{\today}
\begin{abstract}
In this paper we study the formation of topological Tamm states at
the interface between a semi-infinite one-dimensional photonic-crystal
and a metal. We show that when the system is topologically non-trivial
there is a single Tamm state in each of the band-gaps, whereas if
it is topologically trivial the band-gaps host no Tamm states. We
connect the disappearance of the Tamm states with a topological transition
from a topologically non-trivial system to a topologically trivial
one. This topological transition is driven by the modification of
the dielectric functions in the unit cell. Our interpretation is further
supported by an exact mapping between the solutions of Maxwell's equations
and the existence of a tight-binding representation of those solutions.
We show that the tight-binding representation of the 1D photonic crystal,
based on Maxwell's equations, corresponds to a Su-Schrieffer-Heeger-type
model (SSH-model) for each set of pairs of bands. Expanding this representation
near the band edge we show that the system can be described by a Dirac-like
Hamiltonian. It allows one to characterize the topology associated
with the solution of Maxwell's equations via the winding number. In
addition, for the infinite system, we provide an analytical expression
for the photonic bands from which the band-gaps can be computed. 
\end{abstract}
\maketitle

\section{\label{sec:level1}Introduction}

Topology is at the heart of modern condensed matter physics \citep{SZhang:2017,Belopolskie1501692,topology2019,Yan2015}
and photonics \citep{Xie:2018,Rider2019,Ozawa2019,mariosilveirinha}.
It can be found in electronic \citep{murakami2011,Vafek:2014}, photonic
\citep{khanikaev2017,Downing:2017,Bleckmann:2017,Pocock:2018,Whittaker2019,Zhang2019},
acoustic \citep{MengXiao:15,Xin2018,Jiang:2018}, and mechanical \citep{Kane:14}
systems, just to give four examples. Topology in physics refers to
generic electronic properties of a condensed system (or photonic system
if photons are concerned), which are unchanged by continuous deformations
of the Hamiltonian parameters, as long as the gaps in the spectrum
remain open. Such property can be the winding number in one-dimensional
(1D) systems and the Chern number and the $Z_{2}$ invariant in two-dimensional
(2D) ones. Eventually, the continuous change of the Hamiltonian parameters
leads the system to a topological transition where end-states are
allowed in the gaps in the spectrum in the regime where the system
is topologically non-trivial (finite winding or Chern numbers \citep{Watson1996},
for example). This transition requires the closing and reopening of
the gaps in the spectrum at some point of the deformation process.

The so called bulk-edge correspondence \citep{Delplace2011,Rhim2017}
allows to predict, from a bulk property of the system, the existence
of end-states in 1D or edge-states in 2D systems. In a physical system,
such as an electronic one, the edge-states are responsible, for example,
by dissipationless transport \citep{Thouless1982,kohmoto1985} of
electric charge, and the end-states in a 1D photonic crystal are responsible
for a finite transmission coefficient of electromagnetic radiation
in the band-gap of bulk states \citep{Kalozoumis2018}. The connection
of these properties to topological invariants has far reaching consequences,
one of them being the robustness of certain physical properties of
the system making them insensitive to disorder \citep{Meier2018}.

As far as 1D electronic and photonic systems are concerned, the Kronig-Penney
(KP) model is one of the most studied \citep{kronig1931,mcquarrie1996,szmulowicz1997kronig,mishra2003}.
Possibly the first model proposed for understanding the electronic
structure of crystals, it has an infinite number of energy bands and
gaps. Also, in 1D, the Su-Schrieffer-Heeger (SSH) model \citep{Su1980,Su1988},
originally proposed to describe elementary excitations in conducting
polymers, together with its generalizations \citep{Linhu2014,Katsunori2018,Obana2019},
is among the first model known to feature topological behavior and
remains an active field of research till today \citep{Xie2019}. The
SSH model, essentially a tight-binding approximation with energy-dependent
hopping probabilities, is focused on two energy bands. In its simplest
version, the gap depends on the absolute difference of the (constant
but unequal) intra- and inter-cell hopping parameters. The KP and
SSH models, two working horses of electronics and photonics, form
the basis of our understanding of wave propagation in multi-band systems
in 1D. Variants of the latter model have been used to discuss the
formation of end states \citep{glasser1990} and the behavior of light
at interfaces \citep{Istrate_2005,Meng2014}, as well as the role
of defects in dielectric stratified media \citep{liu1997}. Furthermore,
the SSH model has been considered in different contexts, from propagation
of electromagnetic waves in dispersive photonic crystals composed
of meta-materials \citep{wang2009} and topological quantum optics
\citep{Belloeaaw2019} to electronics of artificial condensed matter
structures \citep{Belopolskie1501692}. It has been applied to a variety
of systems, such as topological photonic arrays \citep{El-Ganainy:2015},
light-emitting topological edge states \citep{han2019}, edge states
in a split-ring-resonator chain \citep{Jiang:2018,Xin2018}, and topological
photonic crystal nanocavity lasers \citep{Ota2018}. A recently published
work \citep{Whittaker2019} investigated the role of spin-orbit coupling
on the topological edge modes of a SSH-type model. In the field of
plasmonics the SSH model has also been useful \citep{Gomes:2017,Pocock2019,Jiang2020}.

It has been suggested \citep{sun2017} that there is a certain analogy
between the topological behavior of a 1D photonic crystal and the
topological behavior of the SSH-model \citep{shortTI}. Indeed, in
some conditions, both systems host end-states. In this paper we show
that this analogy is much deeper due to the existence of an exact
mapping between the SSH-model and the simplest 1D-photonic crystal
composed of a unit cell with two different dielectrics. Indeed, using
an appropriate representation of the solutions of Maxwell's equations
for the 1D photonic crystal, equations analogous to the electronic
Kronig-Penny model can be written for the amplitudes of the electric
field at the interface of two dielectrics. However, since the effective
hopping parameters are frequency dependent, it is equivalent to the
SSH-type mode \citep{Su1980}. When the unit cell is homogeneous (same
dielectric function everywhere) the gap closes at the band edge and
the photonic bands disperse linearly and can be described by an effective
Dirac-type Hamiltonian for massless particles. When the unit cell
is inhomogeneous a gap opens at the edge of the Brillouin zone. Using
the tight-binding model obtained from the exact mapping, we shall
discuss the opening and closing of the gap at the band edge using
the Dirac-type Hamiltonian, which allows us to make the connection
to topology via the winding number; this latter topological invariant
can be computed analytically and allows for constructing a kind of
"phase diagram" of the system.

We shall further use the topological point of view considering a special
type of localized photonic state that may arise at the border between
a semi-infinite 1D photonic crystal and another medium, the photonic
Tamm state. It is analogous to the localized electronic Tamm states
predicted to exist at the surface of a crystal owing to the broken
translational symmetry \citep{Tamm1932}. In contrast to the electronic
ones, photonic Tamm states can only exist at the interface between
a periodic photonic structure and a medium with negative dielectric
constant, not at a free surface of the former. The existence of such
states was predicted theoretically \citep{TammPlasmonTP} and later
demonstrated experimentally \citep{sasin_2010} for GaAs/AlGaAs superlattices
covered with a gold layer. The metal can be replaced by another medium
such as a polymer doped with so-called J-aggregates \citep{Sanchez2016}
or a polar crystal with phononic reststrahlen band \citep{Silva2019}.
As we shall see, the existence of photonic Tamm states in the considered
structure can be predicted using topological arguments.

This paper is organized as follows. In Sec. \ref{sec:Photonic-Kronig-Penney-Model},
we introduce the 1D photonic crystal composed of two different dielectrics
in the unit cell and give an approximate analytical expression to
the energy bands (and therefore band gaps) which, to our best knowledge,
cannot be found in the literature so far. In Sec. \ref{sec:Metal-Photonic-Crystal-Interface},
we discuss the problem of the formation of photonic Tamm states at
the interface of a 1D semi-infinite photonic crystal and a metal (modeled
by a complex dielectric permittivity). We identify the existence a
single Tamm state per energy gap. In Sec. \ref{sec:Topology} we make
general considerations on topology of 1D photonic systems which allow
us to appreciate the results of the previous section. In Sec. \ref{sec:Exact-Mapping-to},
we draw an exact mapping between the solution of Maxwell's equations
and a tight-binding model, which exactly coincides with that of the
SSH model. This allows us to formulate the problem of the opening
and closing of the energy gap, as the dielectric functions of the
unit cell vary, in terms of a Dirac-like Hamiltonian. Using this representation,
a connection between the formation of Tamm states and topology is
drew. A conclusions section and three appendices close the paper.

\section{\label{sec:Photonic-Kronig-Penney-Model}Photonic Kronig-Penney Model}

\subsection{Derivation of the dispersion equation}

\begin{figure}[h]
\centering{}\includegraphics[width=8cm]{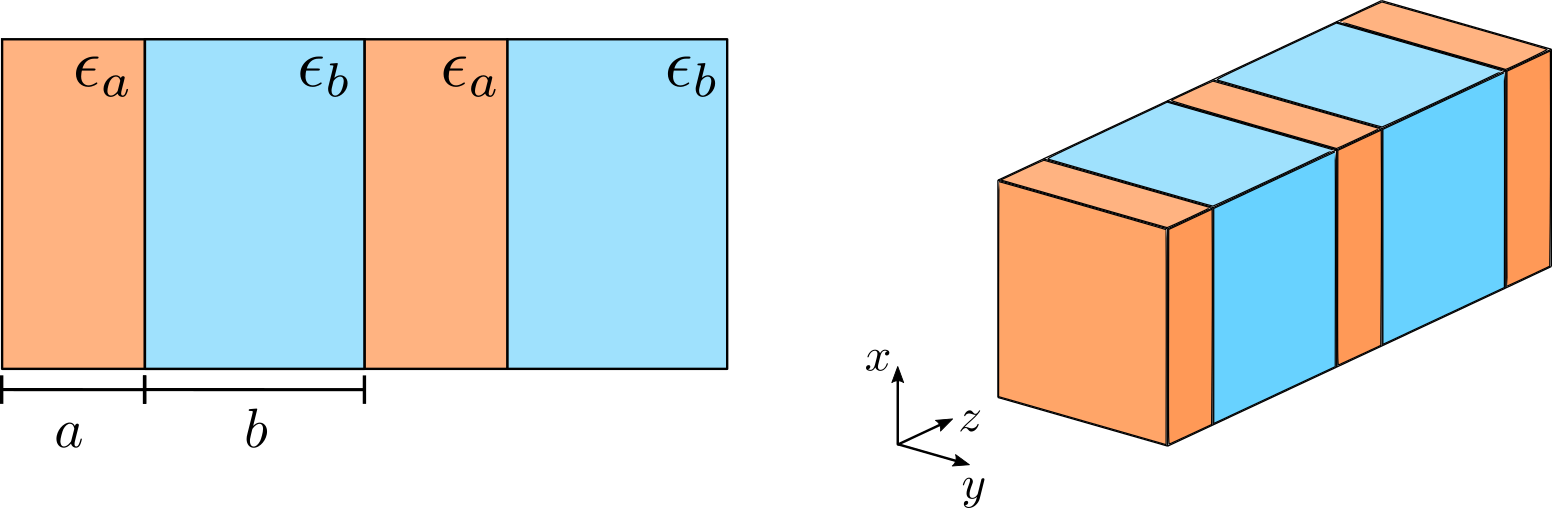}\caption{\label{fig:Infinite PC}Schematic of the considered system: an infinite,
one dimensional, photonic crystal. A side and a perspective view are
presented in the left and right panels respectively. The photonic
crystal is composed of alternating dielectric slabs with dielectric
constant/thickness $\epsilon_{a}$/$a$ and $\epsilon_{b}$/$b$.
Throughout this paper only normal incidence is considered.}
\end{figure}

In this section, we derive the electromagnetic field expressions and
calculate the photonic bands for an infinite photonic crystal (PC).
The representation of the PC is given in Fig. \ref{fig:Infinite PC}.
This section contains both few well-known matter, needed to define
the physical quantities used throughout the paper, and new results,
namely an analytical expression for the photonic bands, $\omega(k)$.

Based on the Maxwell equations, it is possible to formulate electromagnetism
as an eigenvalue problem \citep{joannoupolos,photonicsbook}. For
the magnetic field $\mathbf{H}$, the eigenvalue equation reads: 
\begin{equation}
\mathbf{\boldsymbol{\nabla}}\times\left(\frac{1}{\epsilon(z)}\boldsymbol{\nabla}\times\mathbf{H}(z)\right)=\frac{\omega^{2}}{c^{2}}\mathbf{H}(z),\label{eq:H eigen prob}
\end{equation}
where $c$ is the speed of light in vacuum, $\omega$ is the electromagnetic
field's frequency, and $\epsilon(z)$ is the dielectric function of
the photonic crystal unit cell, defined as: 
\begin{equation}
\epsilon(z)=\begin{cases}
\epsilon_{a,} & 0<z<a\\
\epsilon_{b}, & a<z<a+b
\end{cases}.
\end{equation}
The dielectric function has the periodicity of the photonic crystal,
such that $\epsilon(z+d)=\epsilon(z)$, with $d=a+b$ the period of
the crystal. Considering normal incidence only, we write the magnetic
field as \citep{Romano_2010}: 
\begin{equation}
\mathbf{H}(z)=H_{0}\sqrt{d}h(z)\hat{u}_{y},\label{eq:H field}
\end{equation}
with $H_{0}$ the amplitude of the magnetic field. Note that the factor
$\sqrt{d}$ is introduced by mere convenience, since in this way $h(z)$
has the dimension of inverse of square root of length, which is convenient
for normalization purposes. Inserting (\ref{eq:H field}) into the
eigenvalue equation (\ref{eq:H eigen prob}) yields a solution in
the form: 
\begin{equation}
h_{i}(z)=c_{i1}\cos(k_{i}z)+c_{i2}\sin(k_{i}z),
\end{equation}
where $k_{i}=\sqrt{\epsilon_{i}}\omega/c$ and $i=\{a,b\}$. The electric
field follows from the relation: 
\begin{equation}
\mathbf{E}(z)=\frac{i}{\omega\epsilon_{0}\epsilon_{i}}\boldsymbol{\nabla}\times\mathbf{H}(z),
\end{equation}
with $\epsilon_{0}$ vacuum's dielectric constant. The electric field
can thus be written explicitly, in the particular case we are considering,
as: 
\begin{equation}
\mathbf{E}(z)=-\frac{iH_{0}\sqrt{d}}{c\epsilon_{0}}f(z)\hat{u}_{x},\label{eq:E field}
\end{equation}
with $f(z)$ given by: 
\begin{equation}
f_{i}(z)=\frac{c}{\omega\epsilon_{i}}\frac{dh(z)}{dz}=\frac{1}{\sqrt{\epsilon_{i}}}\left[-c_{i1}\sin(k_{i}z)+c_{i2}\cos(k_{i}z)\right].
\end{equation}
Now, we can make use the transfer matrix method \citep{mora1985,Romano_2010}
to obtain the fields at $z+\Delta z$ from their expressions at $z$,
that is: 
\begin{equation}
\left(\begin{array}{c}
h(z+\Delta z)\\
f(z+\Delta z)
\end{array}\right)=T_{i}(\Delta z)\left(\begin{array}{c}
h(z)\\
f(z)
\end{array}\right),\label{eq:Transfer Matrix 1}
\end{equation}
where $T_{i}(\Delta z)$ is the transfer matrix that shifts the fields
by $\Delta z$ inside a slab with dielectric constant $\epsilon_{i}$,
defined as: 
\begin{equation}
T_{i}(\Delta z)=\left(\begin{matrix}\cos(k_{i}\Delta z) & \sqrt{\epsilon_{i}}\sin(k_{i}\Delta z)\\
-\frac{1}{\sqrt{\epsilon_{i}}}\sin(k_{i}\Delta z) & \cos(k_{i}\Delta z)
\end{matrix}\right).\label{eq:Transfer Matrix 2}
\end{equation}
This matrix is obtained by explicitly writing two systems of equations:
one with $h(z)$ and $f(z)$, and another with $h(z+\Delta z)$ and
$f(z+\Delta z)$. We then use the first system to obtain expressions
for the coefficients $c_{i1}$ and $c_{i2}$. Substituting these into
the second system, we arrive at an expression whose left hand side
is a column vector consisting of the fields $h$ and $f$ evaluated
at $z+\Delta z$ and the right hand side consists of a product of
two matrices and a column vector with $h$ and $f$ evaluated at $z$.
This is equivalent to Eq. (\ref{eq:Transfer Matrix 1}), and the product
of these two matrices gives the transfer matrix $T_{i}(\Delta z)$
(see Appendix \ref{sec:Construction-of-the} for details).

If we now want to move from the origin to the edge of the first unit
cell, we write: 
\begin{equation}
\left(\begin{array}{c}
h(a+b)\\
f(a+b)
\end{array}\right)=T_{a}(a)T_{b}(b)\left(\begin{array}{c}
h(0)\\
f(0)
\end{array}\right).\label{eq:TbTa}
\end{equation}
Invoking Bloch's theorem, this can also be written as: 
\begin{equation}
\left(\begin{array}{c}
h(a+b)\\
f(a+b)
\end{array}\right)=e^{ik(a+b)}\left(\begin{array}{c}
h(0)\\
f(0)
\end{array}\right),\label{eq:Bloch PC}
\end{equation}
where $k$ is Bloch's momentum. We notice that the matrix $T_{a}(a)T_{b}(b)$
is unimodular and, therefore, its eigenvalues are of the form $\lambda$
and $1/\lambda$. Moreover, the trace of a matrix is preserved under
unitary transformations, since $\textrm{Tr}(U^{\dagger}AU)=\textrm{Tr}(AUU^{\dagger})=\textrm{Tr}(A)$.
Thus, we obtain a transcendental equation for the photonic crystal
problem: $2\cos[k(a+b)]=\textrm{Tr}[T_{a}(a)T_{b}(b)]$, which has
the following explicit form: 
\begin{align}
2\cos[k(a+b)] & =2\cos(k_{a}a)\cos(k_{b}b)\nonumber \\
 & -\frac{\epsilon_{a}+\epsilon_{b}}{\sqrt{\epsilon_{a}\epsilon_{b}}}\sin(k_{a}a)\sin(k_{b}b).\label{eq:Transc Photonic}
\end{align}
This equation defines the frequency spectrum of the 1D photonic crystal.
It is a well known result, an implicit relation for $\omega(k)$ (where
$\omega$ enters through $k_{a}$ and $k_{b}$). Even though $k$
can be obtained immediately in terms of $\omega$, analytical inversion
of this relation is not known, so usually Eq. (\ref{eq:Transc Photonic})
is solved numerically.

\subsection{Photonic bands: Analytical results}

Having arrived at Eq. (\ref{eq:Transc Photonic}), we now wish to
solve it in order to obtain the photonic band structure. This equation
can easily be solved numerically, however, we wish to obtain an analytical
expression for $\omega(k)$. As shown below, with a careful choice
of approximations, we can obtain an expression for $\omega(k)$ in
total agreement with the numerical solution. The advantage of this
is an explicit expression for the band in terms of the photonic crystal
parameters, including the calculation of the group velocity, density
of states, and band-gap.

\begin{figure}[h]
\centering{}\includegraphics[width=8cm]{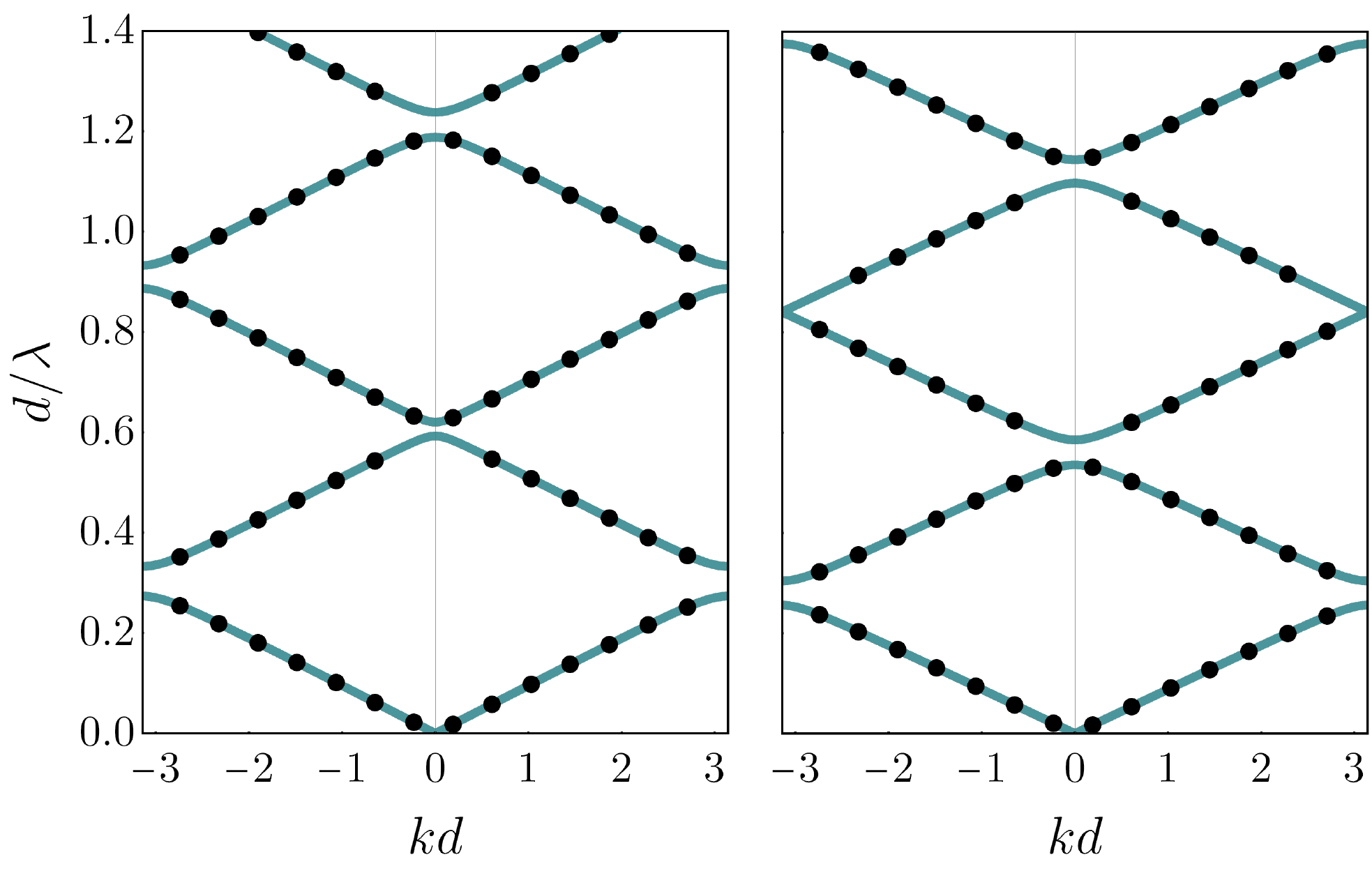} \label{fig:PC bands}
\caption{Plot of the photonic band structure (plotted in terms of dimensionless
variables, $d/\lambda=\omega d/2\pi c$ vs $kd$) for two different
$f=a/(a+b)$ values (left panel $f=0.35$; right panel $f=0.6$) obtained
analytically using Eq. (\ref{eq:Analytical Sol Phot}) (solid lines)
and the exact numerical results (dots). The excellent agreement between
the two approaches is clear. The lowest band was obtained using $n=1$
and $p=0;$ for the second lowest band $n=2$ and $p=0$ were used,
and similarly for the others. To obtain plots in both panels, the
parameters $d=a+b=400$ nm, $\epsilon_{a}=4$ (approximately the dielectric
constant of HfO$_{2}$ in the visible), and $\epsilon_{b}=2.13$ (approximately
the dielectric constant of SiO$_{2}$ in the visible) were considered;
this set of parameters is used throughout the rest of the figures. }
\end{figure}

Let us start by writing the equality: 
\[
(\epsilon_{a}+\epsilon_{b})/\sqrt{\epsilon_{a}\epsilon_{b}}=2+\delta\,,
\]
with $\delta>0$, which is valid for any choice of $\epsilon_{i}$.
Notice that $\delta=0$ occurs when $\epsilon_{a}=\epsilon_{b}$,
that is, this parameter measures the dielectric constant contrast
in the crystal. Inserting this relation into Eq. (\ref{eq:Transc Photonic})
we obtain: 
\begin{equation}
2\cos[k(a+b)]=2\cos(k_{a}a+k_{b}b)-\delta\sin(k_{a}a)\sin(k_{b}b).\label{eq:Photonic Delta}
\end{equation}
This equation can be solved by iterations using $\delta$ as a small
parameter (notice that $\delta=0.01$ corresponds to a substantial
difference of about 20\% between $\epsilon_{a}$ and $\epsilon_{b}$).
Taking the limit $\delta\rightarrow0$, the first approximation for
$\omega$ is easily obtained: 
\begin{eqnarray*}
 & \omega_{0}^{n,p}=\frac{2\pi c}{(\sqrt{\epsilon_{a}}a+\sqrt{\epsilon_{b}}b)}(n-1)-\\
 & (-1)^{n+p}\frac{|k|c(a+b)}{(\sqrt{\epsilon_{a}}a+\sqrt{\epsilon_{b}}b)}\,,
\end{eqnarray*}
with $n=\{1,2,3,...\}$ and $p=\{0,1\}$. Different combinations of
these two indices allow us to describe different bands. While the
index $n$ controls the band's vertical position, the index $p$ controls
its concavity. If $n+p$ is even the band has a bell-like shape; if
$n+p$ is odd the bell shape is turned upside-down. For each value
of $n$ there are two different possible values of $p$. Inserting
this first approximation for $\omega$ in the last term of Eq.(\ref{eq:Photonic Delta}),
produces the solution: 
\begin{eqnarray}
 & \omega^{n,p}(k)=\frac{c}{\sqrt{\epsilon_{a}}a+\sqrt{\epsilon_{b}}b}\times\nonumber \\
 & \left[2\pi(n-1)-(-1)^{n+p}\arccos\psi^{n,p}(k)\right],\label{eq:Analytical Sol Phot}
\end{eqnarray}
where $\psi^{n,p}(k)$ reads: 
\begin{eqnarray}
 & \psi^{n,p}(k)=2\cos[k(a+b)]+\nonumber \\
 & \delta\sin\left(\frac{a\sqrt{\epsilon_{a}}\omega_{0}^{n,p}}{c}\right)\sin\left(\frac{b\sqrt{\epsilon_{b}}\omega_{0}^{n,p}}{c}\right).
\end{eqnarray}
The bands obtained with this expression, as well as the exact results
obtained numerically, are plotted in Fig. \ref{fig:PC bands} where
it is possible to see an excellent agreement between both approaches;
this agreement extends across all the plotted bands. Finally, we note
that, although Eq. (\ref{eq:Analytical Sol Phot}) was obtained after
considering the limit $\delta\rightarrow0$, the analytical solution
holds even when $\epsilon_{a}$ and $\epsilon_{b}$ are significantly
different, as in the case in Fig. \ref{fig:PC bands}.

\begin{figure}[h]
\centering{}\includegraphics[width=8cm]{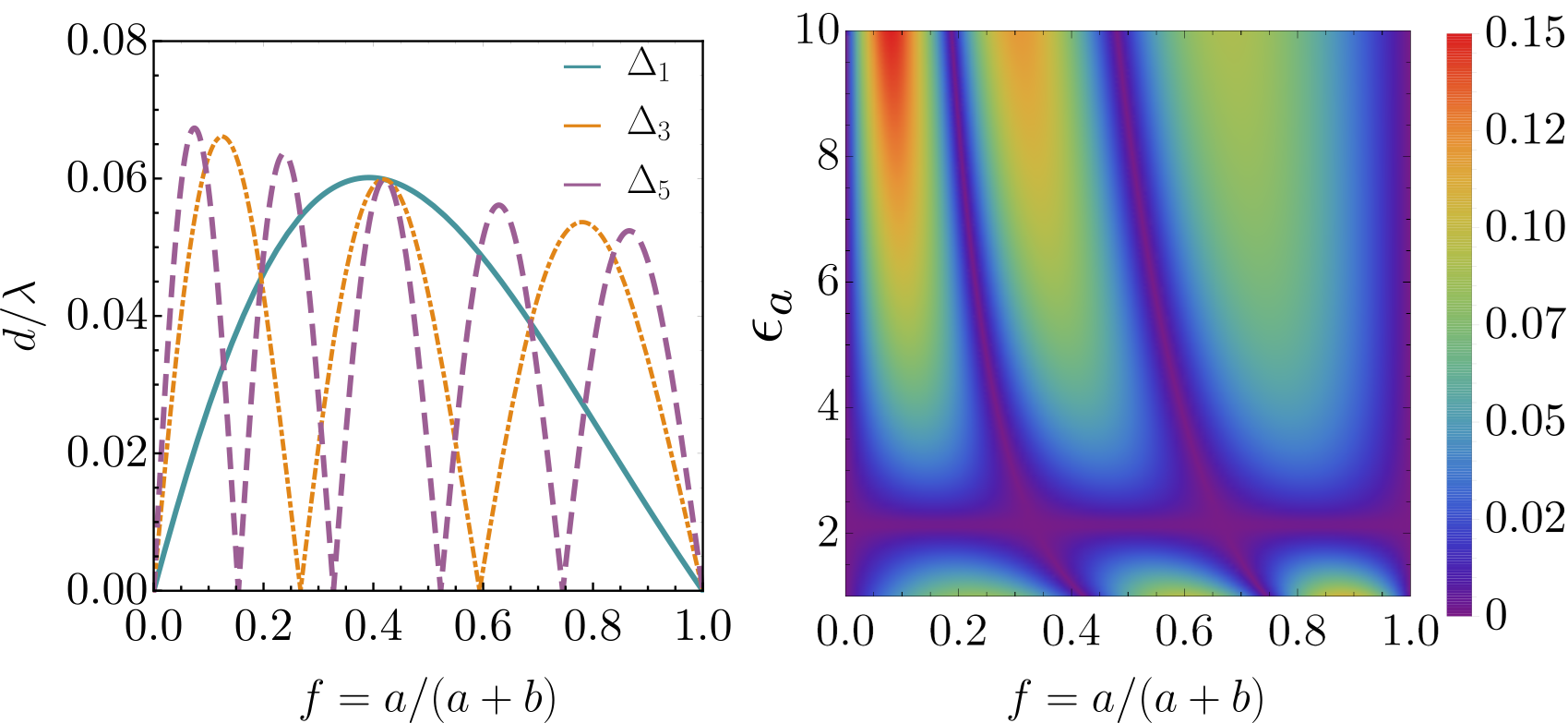}\caption{\label{fig:Gaps}Left panel: Plot of the first, third and fifth gaps
(first three that open at the edge of the Brillouin zone) as a function
of $f=a/(a+b).$ We observe that higher energy gaps close more often
than the ones below. To obtain the left panel the parameters $d=a+b=400$
nm, $\epsilon_{a}=4$ and $\epsilon_{b}=2.13$ (as in Fig. \ref{fig:PC bands})
were considered. Right panel: Color map of the third gap as a function
of both $f$ and $\epsilon_{a}$ for $\epsilon_{b}=2.13$ and $(a+b)=400$
nm. 
}
\end{figure}

One of the main advantages of obtaining analytical expressions for
the bands is the possibility to obtain explicit expressions for the
band gaps that enables one to predict under which circumstances will
the gaps close and reopen. The analytical expression for the gap between
the first two bands reads: 
\begin{align}
\Delta_{1}=\frac{2\hbar c}{a\sqrt{\epsilon_{a}}+b\sqrt{\epsilon_{b}}}\arccos\Bigg[1-\frac{(\sqrt{\epsilon_{a}}-\sqrt{\epsilon_{b}})^{2}}{2\sqrt{\epsilon_{a}\epsilon_{b}}}\cdot\nonumber \\
\cdot\sin\left(\frac{a\pi\sqrt{\epsilon_{a}}}{\sqrt{\epsilon_{a}}a+\sqrt{\epsilon_{b}}b}\right)\sin\left(\frac{b\pi\sqrt{\epsilon_{b}}}{\sqrt{\epsilon_{a}}a+\sqrt{\epsilon_{b}}b}\right)\Bigg]\label{eq:first gap}
\end{align}
In the limit $\epsilon_{a}=\epsilon_{b}$ the band gap vanishes, as
expected, since the system becomes an homogeneous medium. Similarly
to Eq. (\ref{eq:first gap}) we can also obtain analytical expressions
for the other gaps. In the left panel of Fig. \ref{fig:Gaps} we plot
the first, third and fifth gaps, the first three that open at the
edge of the first Brillouin zone, as a function of $f=a/(a+b)$. There,
we can see that all the gaps are closed when $f=0$ and $f=1$, as
expected. Furthermore, the higher energy gaps close more often than
the ones below them as we scan through the interval $f\in[0,1]$.
In the right panel of Fig. \ref{fig:Gaps} we depict the third gap
as a function of $f$ and $\epsilon_{a}$ for $\epsilon_{b}=2.13$.
When either $f=1$ or $\epsilon_{a}=2.13$ the gap closes, which is
consistent with our previous results. Moreover, we observe that the
gap broadens as the difference between the dielectric constants increases
and for $a\ll b$.

\subsection{Determining the fields inside each unit cell}

Let us now focus on finding analytical expressions for the electric
and magnetic fields . To fully describe the fields across the whole
photonic crystal, four coefficients must be determined: $c_{a1}$,
$c_{a2}$, $c_{b1}$, and $c_{b2}$, two for each dielectric slab.
In order to obtain the relations between the different coefficients
we return to the transfer matrix introduced in Eq. (\ref{eq:Transfer Matrix 2}),
written in the form: 
\begin{equation}
T_{b}(b)T_{a}(a)=\left(\begin{matrix}t_{11} & t_{12}\\
t_{21} & t_{22}
\end{matrix}\right),\label{eq: Trans Matrix elements}
\end{equation}
with the entries given in Appendix \ref{sec:Transfer-matrix-elements}.
Recalling Eqs. (\ref{eq:TbTa}) and (\ref{eq:Bloch PC}) and explicitly
writing $h(z)$ and $f(z)$, one obtains the following relation between
the coefficients $c_{a1}$ and $c_{a2}$: 
\begin{equation}
c_{a2}=c_{a1}\sqrt{\epsilon_{a}}\frac{e^{ik(a+b)}-t_{11}}{t_{12}}.\label{eq:ca2 ca1}
\end{equation}
Demanding that the fields must be continuous at $z=a$, two additional
relations appear: 
\begin{align}
c_{b1} & =\cos(k_{a}a)[c_{a1}\cos(k_{a}a)+c_{a2}\sin(k_{a}a)]\nonumber \\
 & +\sqrt{\frac{\epsilon_{b}}{\epsilon_{a}}}\sin(k_{b}a)[c_{a1}\sin(k_{a}a)-c_{a2}\cos(k_{a}a)],
\end{align}

\begin{align}
c_{b1} & =\sqrt{\frac{\epsilon_{b}}{\epsilon_{a}}}\cos(k_{a}a)[c_{a2}\cos(k_{a}a)-c_{a1}\sin(k_{a}a)]\nonumber \\
 & +\sin(k_{b}a)[c_{a1}\cos(k_{a}a)+c_{a2}\sin(k_{a}a)].
\end{align}
The coefficient $c_{a1}$ is determined from the normalization condition.
Therefore, determining $c_{a1}$ is sufficient to entirely describe
the fields inside the first unit cell. In fact, we will note compute
the actual value of $c_{a1}$ and set it equal to 1 when plots of
the fields are presented. To obtain the fields in the rest of the
photonic crystal, Bloch's theorem must be used. If, for example, we
wish to obtain the magnetic field in the second unit cell, we need
only to multiply the magnetic field defined in the first unit cell
by $e^{ik(a+b)}$ while shifting its argument by $(a+b)$, that is
$x\rightarrow x-(a+b)$.

\section{\label{sec:Metal-Photonic-Crystal-Interface}Metal-Photonic Crystal
Interface}

\begin{figure}[h]
\centering{}\includegraphics[width=8cm]{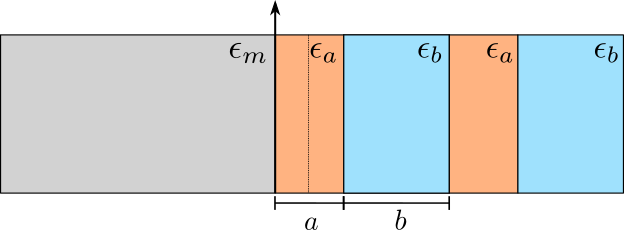}\caption{\label{fig:Semi-Inf PC + Metal}Side view of a semi-infinite photonic
crystal (Bragg mirror) connected to a semi-infinite metal. The semi-infinite
photonic crystal is similar to that of Fig. \ref{fig:Infinite PC}.
The metal slab is also semi-infinite and its dielectric constant is
$\epsilon_{m}$ (note that $\epsilon_{m}$ is complex and will be
exemplified by that of Silver). These two elements of the system are
connected at $z=0$. The dashed line at $z=a/2$ indicates the position
of the inversion center of the infinite crystal. We always consider
an $a$-terminated photonic crystal, where the dielectric function
of the material $a$ is in contact with the metal.}
\end{figure}

Up to this point, we have dealt with an infinite photonic crystal,
determining its band structure as well as the electric and magnetic
fields. In this section, we no longer study the problem of an infinite
photonic crystal, but rather consider a semi-infinite crystal whose
end is connected to a semi-infinite metal, which we will consider
to be Silver. Topologically speaking, the silver is a trivial system.
We are therefore coupling a trivial system to a photonic crystal with
the potential of being topological (this aspect will be addressed
in the next section). This new system is depicted in Fig. \ref{fig:Semi-Inf PC + Metal}.
Our goal is to find the surface states at the metal-photonic crystal
interface. These surface states are usually dubbed photonic Tamm states.

We still consider the magnetic field of the form (\ref{eq:H field})
but this time $h(z)$ is given by: 
\begin{equation}
h(z)=\begin{cases}
Be^{k_{m}z}, & -\infty<z<0\\
c_{a1}\cos(k_{a}z)+c_{a2}\sin(k_{a}z), & 0<z<a
\end{cases},
\end{equation}
with $k_{m}=\sqrt{\epsilon_{m}}\omega/c$, $\epsilon_{m}$ the metal's
dielectric constant and $B$ a coefficient still to be determined.
Note that the argument of the exponential for negative $z$ guarantees
that the field vanishes at large distances. The electric field is
once again obtained using Eq. (\ref{eq:E field}). From the continuity
of $h(z)$ and $f(z)$ at $z=0$, one obtains: $c_{a1}=B$ and $c_{a2}k_{a}/\epsilon_{a}=Bk_{m}/\epsilon_{m}$.
We now recall that we have already obtained in Eq. (\ref{eq:ca2 ca1})
an equality that relates $c_{a2}$ with $c_{a1}$. Substituting this
into the previous two equations, and taking the quotient between them
we arrive at the following relation: 
\begin{equation}
\frac{k_{a}}{\sqrt{\epsilon_{a}}}\frac{e^{ik(a+b)}-t_{11}}{t_{12}}-\frac{k_{m}}{\epsilon_{m}}=0.\label{eq:Metal-PC transc}
\end{equation}
Therefore, we have two coupled equations, Eqs. (\ref{eq:Transc Photonic})
and (\ref{eq:Metal-PC transc}), that need to be solve in order to
obtain the surface states of the metal-photonic crystal interface.
Before doing so, we note that the Bloch momentum of a surface state
has the form \citep{STESLICKA1974,glasser1990} $k=n\pi/d+i\mu$,
where $d=a+b$, $n=\{0,1,2,...\}$ and $\mu$ is a complex number
owing to the imperfect metal. This number $n$ is not related to the
one introduced when the analytical expressions for the infinite photonic
crystal were presented. A state is called even or odd according to
the parity of $n$.

\begin{figure}[h]
\begin{centering}
\includegraphics[width=8cm]{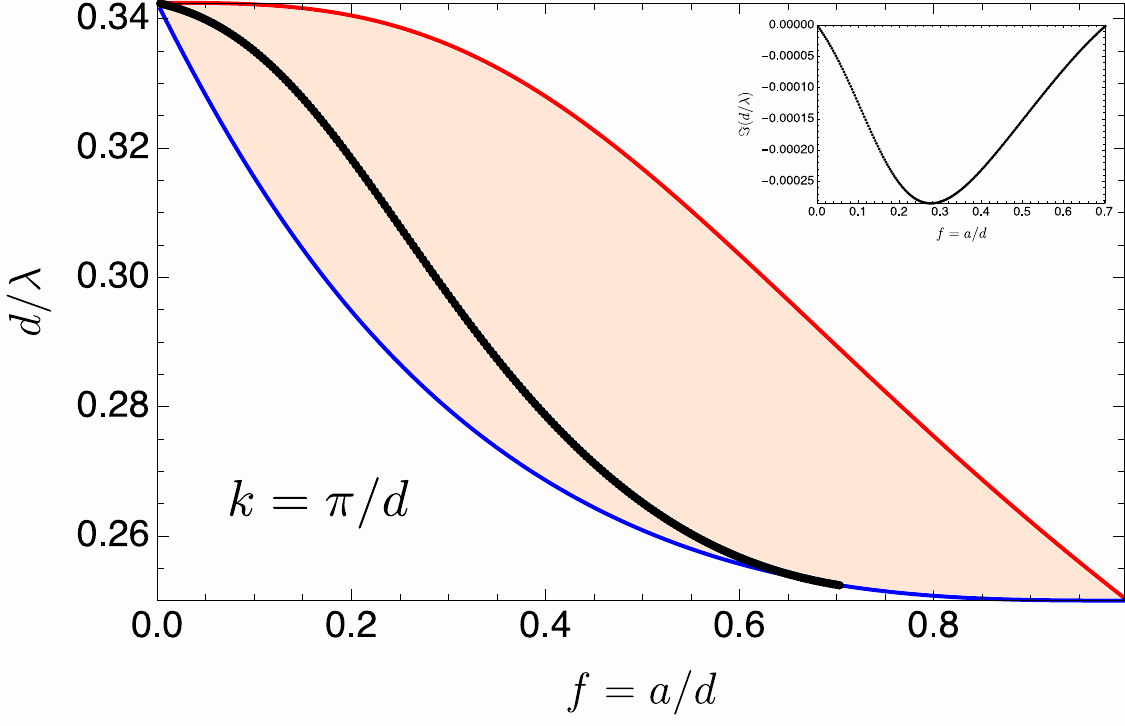} 
\par\end{centering}
\centering{}\includegraphics[width=8cm]{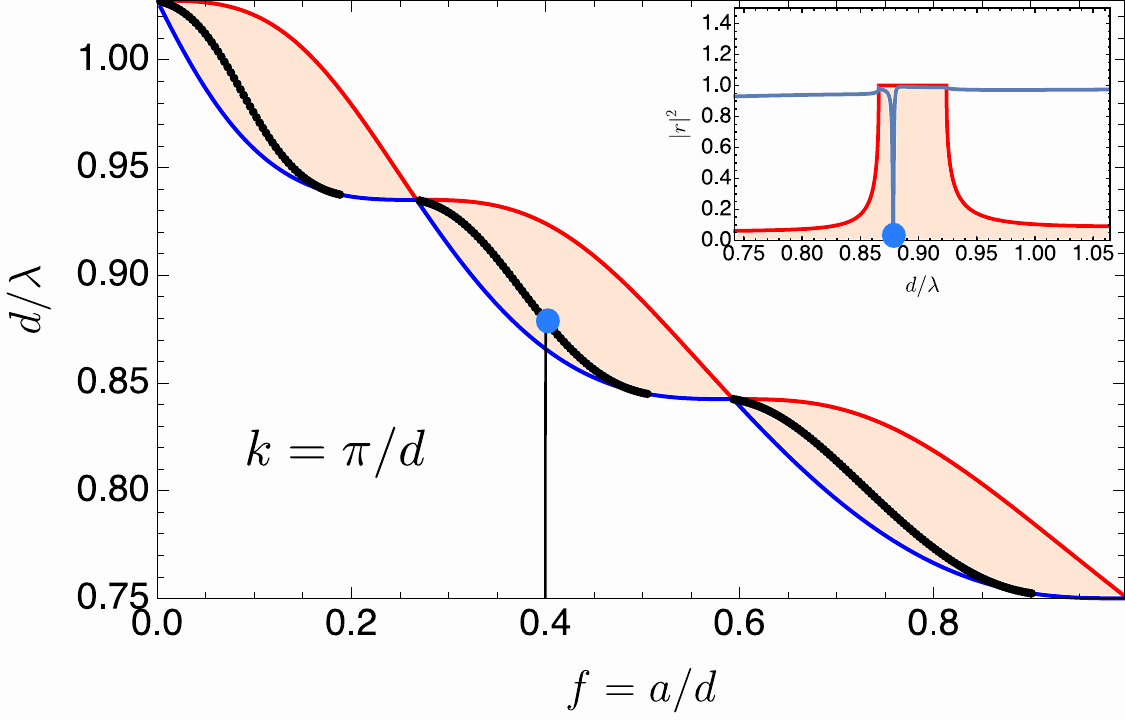}\caption{\label{fig:Gaps and Surf Metal}Representation of the surface state's
energy as a function of $f=a/(a+b)$ (black line). From top to bottom:
first two gaps (shaded regions) of the photonic crystal at the edge
of the Brillouin zone (similar images hold at the center of the zone).
The blue and red lines correspond to the different band edges at $k=\pi/(a+b)$
. When these lines touch each other, the band gap closes. We notice
that the Tamm states (black lines) only appear inside the gaps. Furthermore,
for a given $f$, every gap contains a single surface state. These
graphs are computed using the same PC crystal parameters as in Fig.
\ref{fig:PC bands} and $\epsilon_{m}=-17+i0.5$ (dielectric constant
of Ag at 2 eV). In the inset of the top panel the imaginary part of
$d/\lambda$ is depicted; note that $\Im(d/\lambda)/\Re(d/\lambda)\ll1$,
that is, the effect of dissipation in the metal has a minute effect
in the spectrum of the surface states. In the inset of the bottom
panel we depict the reflectance (blue curve) of a semi-infinite photonic
crystal terminated by a 34 nm Silver film superimposed on the reflectance
(red curve) a purely semi-infinite photonic crystal without the metallic
film. The presence of the Tamm state is clearly seen in the former
(the full dielectric function of Silver was used). }
\end{figure}

Solving Eqs. (\ref{eq:Transc Photonic}) and (\ref{eq:Metal-PC transc})
numerically, we obtain solutions in the form of pairs $(\mu,\omega)$
that characterize the surface states of this system. When solving
these equations numerically, some non-physical solutions may appear.
In order to identify them, some consistency checks must be made: (i)
The value of $\mu$ must be positive, since otherwise the fields would
diverge as $z$ approaches positive infinity, and (ii) $t_{12}$ must
be finite for every ($\mu,\omega$) pair, or else Eq. (\ref{eq:Metal-PC transc})
would not be valid. In agreement with the general expectation, all
the obtained surface states are located inside the band gaps. Furthermore,
we found no more than one state in each gap. In Fig. \ref{fig:Gaps and Surf Metal},
we depict the energy of the surface states as a function of $f=a/(a+b).$
Analyzing this plot we confirm what was previously stated, the surface
states exist only inside the band gaps and there is only one state
per gap for a given $f$. Note that in Fig. \ref{fig:Gaps and Surf Metal}
we have chosen $\epsilon_{a}>\epsilon_{b}$. Had we chosen $\epsilon_{a}<\epsilon_{b}$
and no surface state would exist in the limit we choose a highly conducting
metallic film. This behavior hints at the existence of two different
regimes in the semi-infinite crystal tuned by the value of $\epsilon_{a}$
for $\epsilon_{b}$ fixed (we discuss these two regimes in the context
of the SSH-model in Sec. \ref{sec:Exact-Mapping-to}). Also note that
the Tamm states merge to the band-edge when starting to approach the
closing of a gap. The more negative the metallic film permittivity
is the later the Tamm state merges to the band edge when the closing
of a gap is approaching. Also, in the inset of the third panel of
Fig. \ref{fig:Gaps and Surf Metal}we depict the reflectance (blue
curve) of a semi-infinite photonic crystal terminated by 34 nm Silver
film superimposed on the reflectance (red curve) a purely semi-infinite
photonic crystal without the metallic film. The presence of the Tamm
state is clearly seen in the former (the full dielectric function
of the Silver film was used). The Tamm state is made visible due to
the dissipative dielectric function of Silver and coupling to the
external photonic modes. Also, in the inset of the first panel we
depict the imaginary part of the frequency of the surface-mode existing
in the first gap. We see that this quantity is much smaller than the
real part of the frequency and, therefore, the mode is weakly damped.

\begin{figure}[tbh]
\centering{}\includegraphics[scale=0.6]{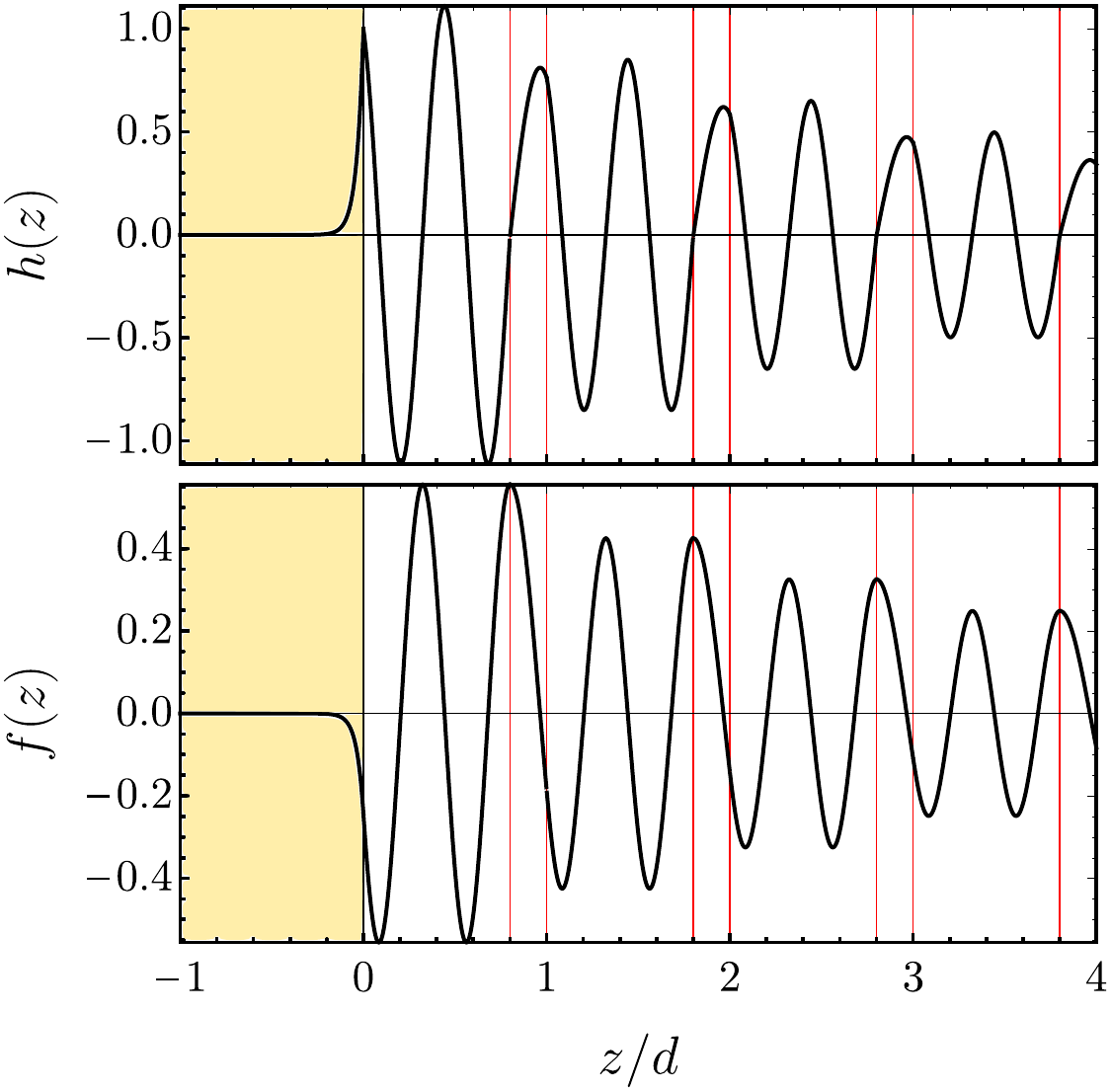}\caption{\label{fig:H E fields Metal}Representation of the real part of $h(z)$
and $f(z)$, which following Eqs. (\ref{eq:H field}) and (\ref{eq:E field})
are proportional to $\mathbf{H}(z)$ and $\mathbf{E}(z)$, respectively,
for a representative Tamm state. This plot is computed using $d=a+b=400$
nm, $\epsilon_{a}=4$ (dielectric constant of HfO$_{2}$), $\epsilon_{b}=2.13$
(dielectric constant of SiO$_{2}$) and $\epsilon_{m}=-17+i0.5$ (dielectric
constant of Ag at 2 eV). This state is obtained for $a/(a+b)=0.8$
and has $\mu d=0.27$ and $d/\lambda=1.05$. In the metal region ($z<0$)
both $h(z)$ and $f(z)$ decay exponentially, as expected (see shaded
region). Inside the photonic crystal ($z>0$) both fields oscillate
while gradually decaying.}
\end{figure}

Next, we compute the electric and magnetic fields of the surface states.
To do so, we follow the formalism presented in the previous section
to construct the fields inside the first unit cell. As was already
discussed, to obtain the field in the rest of the crystal, Bloch's
theorem is invoked. In Fig. \ref{fig:H E fields Metal} we present
the electric and magnetic fields for a representative surface state.
There, we observe that inside the metal both fields decay exponentially
away from the origin; inside the photonic crystal the fields oscillate
while gradually decaying, although at a lower rate than in the metal.
This is characteristic of photonic Tamm states.

\section{\label{sec:Topology}Topological Aspects of the 1D photonic crystal}

Similarly to what occurs in one-dimensional solids, the Bloch functions
of the magnetic and electric fields of a 1D photonic crystal pick
up a Berry phase when $k$ sweeps the Brillouin zone (BZ). If the
system presents inversion symmetry, this phase is quantized and it
is known as Zak's phase, assuming the values of $0$ or $\pi$~\citep{Zak1989}
that can be associated to two distinct topological phases. In a periodic
quantum mechanical system, the Zak's phase, $\gamma_{n}$, for an
isolated band of index $n$ can be expressed as the integration of
the Berry connection $\Gamma_{n,k}$ in the BZ, that is, $\gamma_{n}=\int_{-\pi/d}^{\pi/d}{\Gamma}_{n,k}dk$
where

\begin{equation}
\Gamma_{n,k}=i\int_{0}^{d}dzu_{n,k}^{*}(z)\frac{\partial u_{n,k}(z)}{\partial k}
\end{equation}
and $u_{n,k}(z)$ is the Bloch factor for band $n$. Zak's phase $\gamma$
for a gap above band $n$ is given by the sum $\gamma=\sum_{1}^{n}{\gamma_{n}}$.

One of the simplest examples of 1D systems where one can explore the
transition between the topological phases and their physical consequences
is the SSH-model that describes electrons hopping in a one dimensional
lattice with staggered hopping amplitudes $t_{1}$ (inter-cell) and
$t_{2}$ (intra-cell). In momentum space, the Bloch Hamiltonian reads
${\cal H}=d_{x}(k)\sigma_{x}+d_{y}(k)\sigma_{y}$, where $d_{x}(k)=t_{2}+t_{1}\cos(kd)$,
$d_{y}(k)=t_{1}\sin(kd)$, $\sigma_{x/y}$ is the Pauli matrix $x/y$,
and $d$ is the length of the unit cell. For $t_{1}\neq t_{2}$ the
system presents a gap separating the two bands. The topology of this
gap can be characterized by the Zak phase or, alternatively, by the
\textit{winding number} $w=\gamma/\pi$~\citep{shortTI} that counts
the number of times the vector $\mathbf{d}(k)=(d_{x},d_{y})$ traces
out a closed circle around the origin on the $d_{x}-d_{y}$ plane
when Bloch momentum $k$ sweeps the Brillouin zone. The ratio $t_{1}/t_{2}$
defines the topological character of the system.

Although it is tempting to look for a direct analogy between classical
waves in periodic systems and 1D periodic quantum systems, one should
notice that the formulation of Berry phase and Berry connection in
classical waves is somewhat different from the formulation in electronic
systems. It is important to establish this formulation before moving
into the study of topology in classical waves. A detailed derivation
of the Berry phase for electromagnetic waves, including the possibility
of magneto-electric coupling is described in Ref. \citep{Onoda2006}.
Without magneto-electric coupling, the Berry connection can be written
in terms of the Berry connection of electric and magnetic fields as
$\Gamma_{n,k}=(\Gamma_{n,k}^{E}+\Gamma_{n,k}^{H})/2$ and each contribution
(for the 1D case we are considering in this paper) is given by: 
\begin{align}
\Gamma_{n,k}^{O}=\frac{\int_{0}^{d}dzO_{n,k}^{\ast}(z)\alpha(z)\partial_{k}O_{n,k}(z)}{\int_{0}^{d}dzO_{n,k}^{\ast}(z)\alpha(z)O_{n,k}(z)},\label{eq:Berry}
\end{align}
where $\ast$ represents complex conjugation, $\alpha(z)=\epsilon(z)$
is the (spatial dependent) permittivity for $O_{n,k}(z)=E_{n,k}(z)$,
and $\alpha(z)=\mu(z)$ is the (spatial dependent) permeability for
$O_{n,k}(z)=H_{n,k}(z)$. The fields $E_{n,k}(z)$ and $H_{n,k}(z)$
are the periodic parts of the Bloch wave functions of the $n$th photonic
band, with wave vector $k$, for the electric and magnetic fields,
respectively. Without magneto-electric coupling $\Gamma_{n,k}^{E}=\Gamma_{n,k}^{H}$
and one can choose to use either the electric or the magnetic field
for the Berry connection calculation. In our calculations, as $\mu(z)$
is homogeneous in the PC, one can consider the magnetic field for
the Berry phase calculation.

For a one dimensional system with inversion symmetry, like the SSH-model,
the quantized Zak phase depends on the choice of the inversion center
in the unit cell, that can also be understood as the way one defines
the unit cell containing two different sites. The dependence of the
Zak phase on the specific choice of the inversion center seems to
lead to an ambiguity in the topological aspect of this model. However,
one must recall the bulk-edge correspondence in topological systems.
In connection with bulk properties of this model, there are edge states
for $w=1$. Although there are two possible ways of obtaining $w=1$
for the SSH model that depend on the choice of the inversion center,
there are also two different ways of cutting the chain. If the chain
is cut at the boundary of the unit cell, for example, then for $t_{2}/t_{1}<1$
we have $w=1$ and the finite SSH chain presents localized edge states.
However, the unit cell could be cut in $d/2$, which is equivalent
to redefining the unit cell with a shift of $d/2$. In this case,
the intercell $t_{2}$ hopping becomes the intra-cell one and vice
versa. Therefore, the system presents $w=1$ and localized edge states
for $t_{2}/t_{1}>1$.

In the case of a continuum system such as a photonic crystal, it is
possible to produce an edge in an arbitrary part of the unit cell.
If one wants to discuss the topology and calculate the Berry phase
of the corresponding bulk system, it is necessary to redefine the
origin of the unit cell in the location of the cut, similarly to what
is done in the SSH model for different types of edges. To exemplify
this point, let us consider the semi-infinite PC of Fig. \ref{fig:Semi-Inf PC + Metal}:
the boundary with the metal is not located at any of the inversion
points (one of those signaled by a vertical dashed line) of the bulk
crystal and the crystal's unit cell does not present inversion symmetry.
In this case, the possible Berry phases when sweeping the BZ are not
quantized~\citep{Zak1989}. However, it is still possible to obtain
the Berry phases for arbitrary definitions of the unit cell in terms
of the Zak phases for the inversion symmetric unit cells. The non-quantized
Berry phases still dictate the topological properties of the system
and the existence of edge states in the interface of the PC with other
materials. To calculate the Berry phase for the arbitrary unit cell,
we first notice that the origin of the unit cell is at distance $a/2$
from the closest inversion center (dashed line in Fig. \ref{fig:Semi-Inf PC + Metal}).
Let us consider an inversion symmetric unit cell and perform a translation
by a distance $a/2$: $z^{\prime}=z-a/2$. The Bloch functions have
to be redefined as $u_{k}^{\prime}(z^{\prime})=u_{k}(z-a/2)e^{-ika/2}$.
If the new Bloch factors are used in the calculation of the Berry
connection of Eq.~(\ref{eq:Berry}), one gets ${\Gamma^{\prime}}_{n,k}={\Gamma}_{n,k}+a/2$.
After integrating the new Berry connection in the BZ, the Berry phase
for an arbitrary definition of the unit cell is given by~\citep{Atala2013}
($G$ the smallest reciprocal lattice vector): $\gamma{}_{n}^{\prime}=\gamma_{n}+Ga/2=\gamma_{n}+\pi a/d$.
This gives two non-quantized Berry phases $\gamma_{+}=\pi a/d$ and
$\gamma_{-}=\pi+\pi a/d$ that are related to each other by a $\pi$
phase difference: $\gamma_{-}=\gamma_{+}+\pi$. Although the difference
between the two phases is fixed, their sum varies continuously when
varying the ratio $a/d$, giving rise to a rich phenomenology in multi-band
systems.

In particular, for a metal-PC interface, the existence of edge states
is determined by the sum in the reflection phases of the metal ($\phi_{m}$)
embedded in a medium of dielectric function $\epsilon_{a}$ and the
PC ($\phi_{PC}$); the condition for the existence of an edge state
reads: $\phi_{m}+\phi_{PC}=0$~\citep{Silva2019,Wang2016}. However,
$\phi_{PC}$ is directly related to the Berry phase of the bulk photonic
crystal and consequently it is a function of $a/d$. This variation
gives rise to one edge state per individual gap in Fig. \ref{fig:Gaps and Surf Metal}.

\section{\label{sec:Exact-Mapping-to}Mapping to the SSH-model and the Dirac-like
Hamiltonian}

\begin{figure}[tbh]
\begin{centering}
\includegraphics[width=8cm]{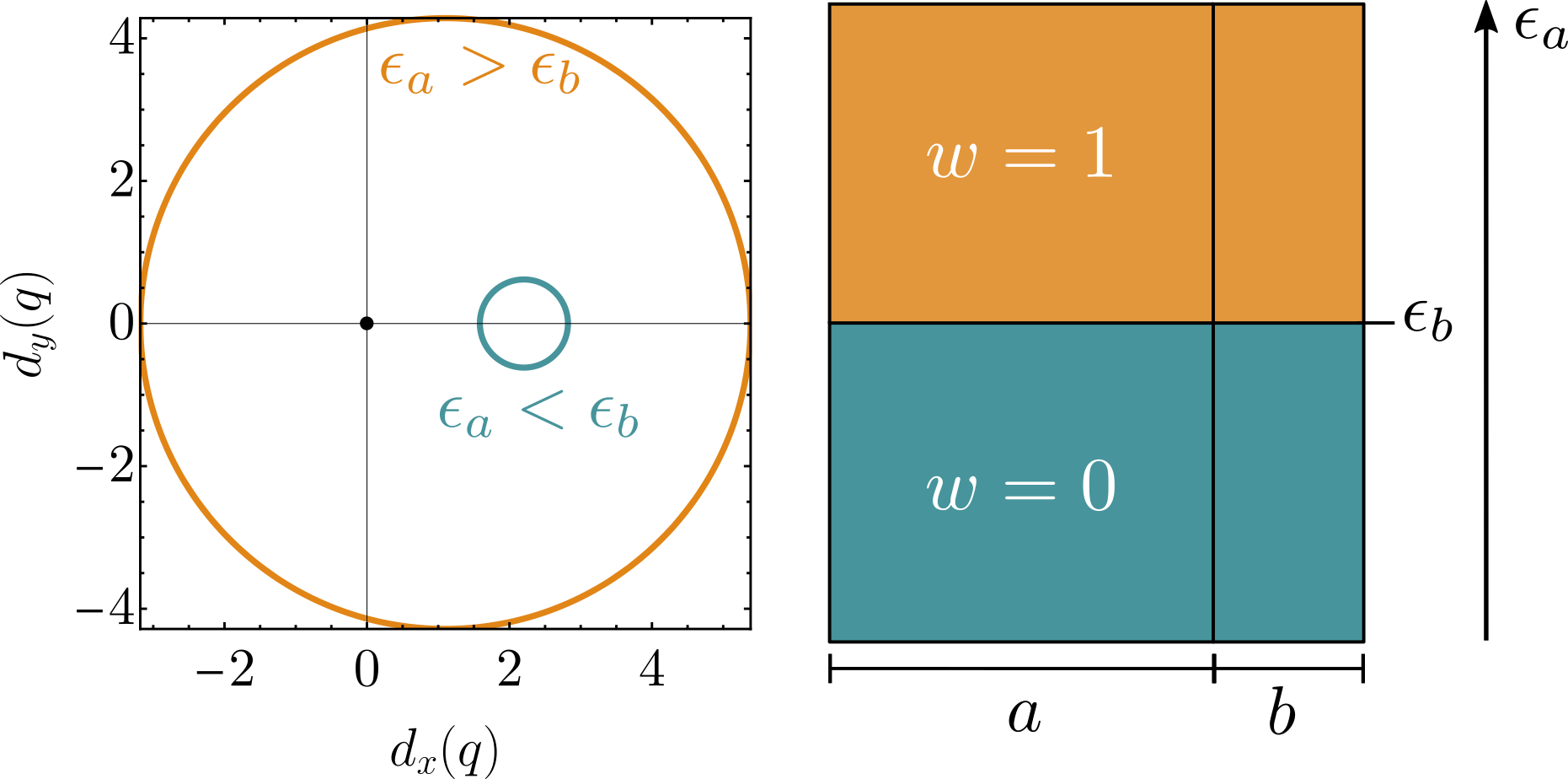} 
\par\end{centering}
\caption{Left panel: Parametric plot of the vector $\mathbf{d}$ illustrating
the topologically non-trivial ($\epsilon_{a}>\epsilon_{b}$) and topologically
trivial ($\epsilon_{a}<\epsilon_{b}$) cases. In the first case the
closed curve includes the origin, whereas in the second case it does
not. Right panel: phase diagram of the photonic SSH-model. The horizontal
axis represents the regime we are considering, $b/a\ll1$, and the
vertical scale shows the increase of $\epsilon_{a}$ compared to a
fixed value of $\epsilon_{b}$. When $\epsilon_{a}>\epsilon_{b}$
the winding number $w$ jumps from $0$ to $1$. We note that the
same conclusions hold even if we use the full frequency dependence
of the hopping parameters except for the shape of the curves that
are no longer circumferences. \label{fig:Parametric-plot-of}}
\end{figure}

In Sec. \ref{sec:Metal-Photonic-Crystal-Interface} we studied a semi-infinite
photonic crystal connected to a semi-infinite metal slab. In that
configuration we observed the existence of a single Tamm state in
each gap, when $\epsilon_{a}>\epsilon_{b}$, and the absence of that
state in the opposite regime. Here we make the connection of those
results to the SSH-model. We show below that an exact mapping exists
between the solutions of Maxwell's equations of the 1D photonic crystal
and the SSH tight-binding model. Once this mapping is established,
we show that a Dirac-like Hamiltonian can be written around the band
edge thus allowing the connection between the results of Sec. \ref{sec:Metal-Photonic-Crystal-Interface}
and topology. Following the procedure described in detail in Appendix
\ref{sec:Derivation-of-the} we can show that the solutions of Maxwell's
equations obey, in real space, to the following set of two equations
(this result is exact):

\begin{align}
-A\phi_{n}+C\psi_{n}-B\phi_{n+1} & =0,\\
-A\psi_{n}+C\phi_{n}-B\psi_{n-1} & =0.
\end{align}
which we instantly recognize as a set of tight-binding equations identical
to those of the SSH-model \citep{shortTI}, and where the different
parameters, $A$, $B$, and $C$ entering tight-binding equations
are defined as in Eqs. (\ref{eq:A})-(\ref{eq:phi_n}). Clearly, the
parameter $A$ is an intra-unit-cell hopping and $B$ is an inter-unit-cell
one. The parameter $C$ represents an onsite energy. We should remark,
however, that contrary the usual tight-binding parameters in the SSH-model,
the parameters $A$, $B$, and $C$ are energy dependent. This is
essential for obtaining the spectrum condition (\ref{eq:Transc Photonic})
from the diagonalization of the tight-binding equations. However,
since the topological nature of the system is determined by the opening
and closing of the gap (leading to band inversion) we can focus our
attention in a single gap at the time. Therefore we can expand the
tight-binding parameters around the energy at which the gap closes,
near $q=\pi/d$ (the same arguments apply when the gap closes at $q=0$).
To be specific, we choose the regime where all gaps close, that is
when $f\rightarrow1$ ($b/a\ll1$), as shown in Fig. \ref{fig:Gaps}.
Let us focus our attention in the first gap. It closes at a frequency
given by $\omega_{c}=c\pi/(\sqrt{\epsilon_{a}}d)$ when $\epsilon_{a}=\epsilon_{b}$.
We can therefore expand the tight-binding parameters around this frequency.
This leads to energy-independent hopping parameters $A$ and $B$
in the form 
\begin{align}
A & \approx\frac{1}{\sqrt{\epsilon_{a}}\sin(a\pi/d)},\\
B & \approx\frac{1}{\sqrt{\epsilon_{b}}\sin(b\pi\sqrt{\epsilon_{b}}/(d\sqrt{\epsilon_{a}}))},
\end{align}
and with the onsite energy expanded as

\begin{equation}
C\approx C_{0}+C_{1}(\omega-\omega_{c}),
\end{equation}
with 
\begin{equation}
C_{0}=\frac{1}{\sqrt{\epsilon_{a}}}\cot(a\pi/d)+\frac{1}{\sqrt{\epsilon_{b}}}\cot(b\pi\sqrt{\epsilon_{b}}/(d\sqrt{\epsilon_{a}})),
\end{equation}
and 
\begin{equation}
C_{1}=\frac{b+a+a\cot^{2}(a\pi/d)+b\cot^{2}(b\pi\sqrt{\epsilon_{b}}/(d\sqrt{\epsilon_{a}}))}{c}.
\end{equation}
With these relations we arrive at a tight-binding eigenvalue problem
of the form: 
\begin{align}
A\phi_{n}-C_{0}\psi_{n}+B\phi_{n+1} & =C_{1}\psi_{n}(\omega-\omega_{c}),\\
A\psi_{n}-C_{0}\phi_{n}+B\psi_{n-1} & =C_{1}\phi_{n}(\omega-\omega_{c}),
\end{align}
which maintains its original SSH-model form, but now with energy-independent
parameters. Introducing $\psi_{n}=\psi_{0}e^{iqdn}$ and $\varphi_{n}=\varphi_{0}e^{iqdn}$
we obtain: 
\begin{equation}
\left(\begin{matrix}-C_{0}-C_{1}(\omega-\omega_{c}) & A+Be^{iqd}\\
A+Be^{-ikd} & -C_{0}-C_{1}(\omega-\omega_{c})
\end{matrix}\right)\left(\begin{matrix}\psi_{0}\\
\phi_{0}
\end{matrix}\right)=0.\label{eq:TB_q_space}
\end{equation}
For having non-trivial solutions for the energy bands the determinant
of the previous matrix must be zero, which leads to the following
two-band spectrum: $\omega=\omega_{c}-C_{0}/C_{1}\pm g(k)/C_{1}$,
where $g(k)=\sqrt{A^{2}+B^{2}+2AB\cos(qd)}$. This expression for
the bands holds near $k=\pi/d$. Considering the eigenvalue problem
defined by Eq. (\ref{eq:TB_q_space}), we notice that a Dirac-like
Hamiltonian can be introduced in the form: 
\begin{align}
{\cal H} & =-C_{0}\mathbf{1}+d_{x}(q)\sigma_{x}+d_{y}(q)\sigma_{y},\nonumber \\
 & =-C_{0}\mathbf{1}+\boldsymbol{\sigma}\cdot\mathbf{d}(q)\label{eq:H_Dirac}
\end{align}
with $\mathbf{d}=(d_{x,}d_{y},0)$, and $d_{x}(q)=A+B\cos(qd)$, $d_{y}(q)=-B\sin(qd)$,
and $\bm{\sigma}=(\sigma_{x},\sigma_{y})$ with $\sigma_{x/y}$ the
$x/y$ Pauli-matrix. This Hamiltonian is valid near the band gap at
$q\approx\pi/d$ and for $b\ll a$, and satisfies the eigenvalue equation
${\cal H}(\psi_{0},\phi_{0})^{T}=C_{1}(\omega-\omega_{c})(\psi_{0},\phi_{0})^{T}$,
where $T$ stands for the transposition operation. Equation (\ref{eq:H_Dirac})
is one of the important results of this paper and allows for the discussion
to the topological nature of the end states found in Sec. \ref{sec:Metal-Photonic-Crystal-Interface}.
Indeed, it can be shown that the winding number (a topological invariant)
is given by \citep{shortTI}

\begin{equation}
w=\frac{1}{2\pi}\int_{-\pi/d}^{\pi/d}dq\left(\hat{\mathbf{d}}(q)\times\frac{d\hat{\mathbf{d}}(q)}{dq}\right)_{z},
\end{equation}
where $\hat{\mathbf{d}}(q)=\mathbf{d}/\vert\mathbf{d}\vert$. It is
not difficult to show by direct integration that $w=1$ when the system
hosts Tamm states and $w=0$ when it does not. Therefore, the Tamm
states we have found in Sec. \ref{sec:Metal-Photonic-Crystal-Interface}
are indeed of topological nature. The finiteness of the winding number
is best seen in a parametric plot of the vector $\mathbf{d}(q)$.
The system is topological when the parametric curve encloses the origin
and is trivial when it does not. We represent this behavior in Fig.
\ref{fig:Parametric-plot-of} for one example of a topologically non-trivial
and a topologically trivial cases. The transition from the topological
phase to the trivial phase is controlled, for $d$ and $a$ fixed,
by the relative values of the dielectric constants $\epsilon_{a}$
and $\epsilon_{b}$. Indeed, if $\epsilon_{a}>\epsilon_{b}$ the system
is topologically non-trivial, whereas in the opposite regime the system
is topologically trivial; this transition is depicted graphically
in the right panel of Fig. \ref{fig:Parametric-plot-of}, as a kind
of phase diagram.

\section{Conclusions }

With this work we studied two distinct photonic systems: (i) an infinite
photonic crystal and (ii) a semi-infinite PC connected to a semi-infinite
metal. For the system (i) we used the transfer matrix method to obtain
a transcendental equation that defines the photonic spectrum. Then,
with a careful choice of approximations, we obtained analytical expressions
for the photonic bands, $\omega(k)$, and the band gaps, two results
not known in the literature two our best knowledge. The analytical
results proved to be in excellent agreement with the exact ones obtained
numerically, even for large dielectric contrast.

Afterwards, the problem (ii) was tackled with the objective of finding
the photonic Tamm states. Imposing boundary conditions for the electric
and magnetic fields at the PC surface and combining them with the
transcendental equation previously obtained for the case (i), we derived
a system of equations whose solution gives the spectrum of the Tamm
states. Plotting the frequencies of these states \textit{versus} $f=a/(a+b)$,
we found that every band gap contained exactly one Tamm state when
$\epsilon_{a}>\epsilon_{b}$ and none in the opposite case. Had we
defined a unit cell with inversion symmetry and cut the crystal at
$a/2$ (inversion center), the contact with metal should generate
no Tamm states. A simple reasoning for this can be given based on
the method of images suitable for a perfect metal, which implies that
the semi-infinite crystal reflected with respect to the metal surface,
together with the real semi-infinite crystal would form an infinite
periodic system without any defect at $z=0$ \citep{joannoupolos}.

Further analysis of Maxwell equations' solutions allowed us to introduce
a tight-binding-type Hamiltonian that is a variant of the SSH model
and to analyze it from the topology point of view. It is probably
the central result of this work. From the SSH model, a Dirac-like
Hamiltonian representation of the bulk states near the band edge,
where the band gap closes, was introduced. With this representation
at hand, the winding number was computed and two topologically different
situations have been distinguished: (1) a topologically non-trivial
phase for $\epsilon_{a}>\epsilon_{b}$, and (2) a topologically trivial
one in the opposite case. Although we have cast the analysis of the
topological nature of the Tamm states in terms of the winding number,
we could also have analyzed the problem in terms of the Zak phase
\citep{Zak1989,Gao:2017,vanderbilt,Wang_2019} of the bands, which
can be accessed via reflection measurements \citep{Gao:15}. Indeed,
we can compute the Zak phase for the first two bands and find $\gamma_{1}=0$
and $\gamma_{2}=\pi$ when $\epsilon_{a}>\epsilon_{b}$ and $b/a\ll1$,
and $\gamma_{1}=\gamma_{2}=\pi$ when $\epsilon_{a}<\epsilon_{b}$
and $b/a\ll1$ if choosing a mirror symmetric unit cell (see discussion
in Sec. \ref{sec:Topology}). These results mean that in the first
case the gap has a topological nature, whereas in the second regime
it is a topologically trivial gap (same Zak phase in both bands).

We did not include the effect of dissipation in the photonic crystal
(although we did that for the metal) for two reasons. Firstly, the
materials we are using in the model, HfO$_{2}$ and SiO$_{2}$, have
negligible dissipation in the visible spectral range. Secondly, the
main effect of taking it into account would be adding a small imaginary
part to either wavevector or frequency but no significant changes
in the spectrum should be expected as long as $\Re\epsilon_{a/b}\gg\Im\epsilon_{a/b}$
(see, however, Ref. \citep{Asger2018} for a detailed study of the
role of dissipation in photonic systems). The effect of damping in
the metal (imaginary part of the dielectric function of Silver) is
best seen in the reflectance of the finite metal film on top of the
semi-infinite photonic crystal (results shown in the inset of Fig.
\ref{fig:Gaps and Surf Metal}). The damping makes the Tamm state
clearly visible as a dip in the reflectance spectrum. One possible
way to overcome damping is to use a ${\cal PT-}$symmetric system
where the losses are compensated by the gain \citep{Mortensen:18}.

Finally, we have not studied the case of a finite transverse momentum.
This would make de Tamm states dispersive along the transverse direction,
with some "effective mass" \citep{TammPlasmonTP,Silv2019}. Its
detailed analysis will be the focus of a forthcoming publication.
\begin{acknowledgments}
N.M.R.P., M.I.V., and Y.V.B. acknowledge support from the European
Commission through the project ``Graphene-Driven Revolutions in ICT
and Beyond'' (Ref. No. 785219) and the Portuguese Foundation for
Science and Technology (FCT) in the framework of the Strategic Financing
UID/FIS/04650/2019. N.M.R.P., T.G.R., and Y.V.B. acknowledge COMPETE2020,
PORTUGAL2020, FEDER and the Portuguese Foundation for Science and
Technology (FCT) through project and POCI-01-0145-FEDER-028114. The
authors acknowledge André Chaves for suggesting the starting point
of the analytical approach to the photonic bands. N.M.R.P. acknowledges
stimulating discussions with Joaquin Fernandéz-Rossier on the topic
of the paper. J.C.G.H. acknowledges the hospitality of the Physics
Department of SDU, Denmark, where this work was completed. The authors
are thankful to Asger Mortensen and Mário Silveirinha for their careful
and critical reading of the manuscript. 
\end{acknowledgments}

\appendix

\section{\label{sec:Construction-of-the}Derivation of the transfer matrix}

Within a slab of constant $\epsilon_{i}$ we have the following relation
for the fields and the expansion coefficients $c_{i1}$ and $c_{i2}$:
\begin{widetext}
\begin{equation}
\left[\begin{array}{c}
h(z+\Delta z)\\
f(z+\Delta z)
\end{array}\right]=\left[\begin{array}{cc}
\cos[\alpha_{i}(z+\Delta z)] & \sin[\alpha_{i}(z+\Delta z)]\\
-\frac{1}{\sqrt{\epsilon_{i}}}\sin[\alpha_{i}(z+\Delta z)] & \frac{1}{\sqrt{\epsilon_{i}}}\cos[\alpha_{i}(z+\Delta z)]
\end{array}\right]\left[\begin{array}{c}
c_{i1}\\
c_{i2}
\end{array}\right]
\end{equation}
\end{widetext}

and 
\begin{equation}
\left[\begin{array}{c}
h(z)\\
f(z)
\end{array}\right]=\left[\begin{array}{cc}
\cos(\alpha_{i}z) & \sin(\alpha_{i}z)\\
-\frac{1}{\sqrt{\epsilon_{i}}}\sin(\alpha_{i}z) & \frac{1}{\sqrt{\epsilon_{i}}}\cos(\alpha_{i}z)
\end{array}\right]\left[\begin{array}{c}
c_{i1}\\
c_{i2}
\end{array}\right]\label{eq:c1_and_c2_matrix}
\end{equation}
From the previous two equations we can eliminate the vector composed
of the coefficients $c_{i1}$ and $c_{i2}$. Inverting the last equation
follows 
\begin{equation}
\left[\begin{array}{c}
c_{i1}\\
c_{i2}
\end{array}\right]=\left[\begin{array}{cc}
\cos(\alpha_{i}z) & -\sqrt{\epsilon_{i}}\sin(\alpha_{i}z)\\
\sin(\alpha_{i}z) & \sqrt{\epsilon_{i}}\cos(\alpha_{i}z)
\end{array}\right]\left[\begin{array}{c}
h(z)\\
f(z)
\end{array}\right],
\end{equation}
which leads to 
\begin{equation}
\left[\begin{array}{c}
h(z+\Delta z)\\
f(z+\Delta z)
\end{array}\right]=\left[\begin{array}{cc}
\cos(\alpha_{i}\Delta z) & \sqrt{\epsilon_{i}}\sin(\alpha_{i}\Delta z)\\
-\frac{1}{\sqrt{\epsilon_{i}}}\sin(\alpha_{i}\Delta z) & \cos(\alpha_{i}\Delta z)
\end{array}\right]\left[\begin{array}{c}
h(z)\\
f(z)
\end{array}\right],
\end{equation}
which can be written as 
\begin{equation}
\left[\begin{array}{c}
h(z+\Delta z)\\
f(z+\Delta z)
\end{array}\right]=T_{i}(\Delta z)\left[\begin{array}{c}
h(z)\\
f(z)
\end{array}\right],
\end{equation}
where $T_{i}(\Delta z)$ is the transfer matrix within the region
of constant $\epsilon_{i}$. If we now want to connect two regions
of different $\epsilon_{i}$, where the one to the left is $\epsilon_{a}$
and the subsequent one is $\epsilon_{b}$, we must have 
\begin{equation}
\left[\begin{array}{c}
h(a+b)\\
f(a+b)
\end{array}\right]=T_{b}(b)T_{a}(a)\left[\begin{array}{c}
h(0)\\
f(0)
\end{array}\right].
\end{equation}
The product $T_{b}(b)T_{a}(a)$ defines the transfer matrix across
the unit cell.

\section{\label{sec:Transfer-matrix-elements}Transfer matrix elements}

In this Appendix we briefly present the expressions for the different
entries of the transfer matrix given in Eq. (\ref{eq: Trans Matrix elements}):
\begin{align}
t_{11} & =\cos(k_{a}a)\cos(k_{b}b)-\sqrt{\frac{\epsilon_{b}}{\epsilon_{a}}}\sin(k_{a}a)\sin(k_{b}b),\\
t_{12} & =\sqrt{\epsilon_{a}}\sin(k_{a}a)\cos(k_{b}b)+\sqrt{\epsilon_{b}}\cos(k_{a}a)\sin(k_{b}b),\\
t_{21} & =-\frac{1}{\sqrt{\epsilon_{a}}}\sin(k_{a}a)\cos(k_{b}b)-\frac{1}{\sqrt{\epsilon_{b}}}\cos(k_{a}a)\sin(k_{b}b),\\
t_{22} & =\cos(k_{a}a)\cos(k_{b}b)-\sqrt{\frac{\epsilon_{a}}{\epsilon_{b}}}\sin(k_{a}a)\sin(k_{b}b).
\end{align}

\section{\label{sec:Derivation-of-the}Derivation of the tight-binding electromagnetic
model}

Here we give the details of the derivation of the tight-binding electromagnetic
model, showing the existence of an exact mapping between the solutions
of Maxwell's equations and the SSH model. To do so, we recall the
definition of the electric and magnetic fields presented in the main
text: 
\begin{equation}
\mathbf{H}(z)=H_{0}\sqrt{d}h(z)\hat{u}_{y},
\end{equation}
with 
\begin{equation}
h_{i}(z)=c_{i1}\cos(k_{i}z)+c_{i2}\sin(k_{i}z),
\end{equation}
where $k_{i}=\sqrt{\epsilon_{i}}\omega/c$; and: 
\begin{equation}
\mathbf{E}(z)=\frac{i}{\omega\epsilon_{0}\epsilon_{i}}\mathbf{\mathbf{\nabla}\times\mathbf{H}}(z),
\end{equation}
which written explicitly reads: 
\begin{equation}
\mathbf{E}(z)=-\frac{iH_{0}\sqrt{d}}{c\epsilon_{0}}f(z)\hat{u}_{x},
\end{equation}
with 
\begin{equation}
f(z)=\frac{c}{\omega\epsilon_{i}}\frac{dh(z)}{dz}=\frac{1}{\sqrt{\epsilon_{i}}}[-c_{i1}\sin(k_{i}z)+c_{i2}\cos(k_{i}z)].
\end{equation}

Our goal is to obtain $h(z)$ and $f(z)$ from their values at the
interfaces between dielectric slabs. To this end, we use the transfer
matrix presented in Eq. (\ref{eq:Transfer Matrix 2}) to write: 
\begin{equation}
\left(\begin{matrix}h_{a}(nd+a)\\
f_{a}(nd+a)
\end{matrix}\right)=T_{a}(a)\left(\begin{matrix}h_{a}(nd)\\
f_{a}(nd)
\end{matrix}\right),
\end{equation}
which gives us $h$ and $f$ on the interface between slabs $a$ and
$b$ of the $n-$th unit cell from their values at the origin of the
$n-$th unit cell; and: 
\begin{equation}
\left(\begin{matrix}h_{a}(z)\\
f_{a}(z)
\end{matrix}\right)=T_{a}(z-nd)\left(\begin{matrix}h_{a}(nd)\\
f_{a}(nd)
\end{matrix}\right),
\end{equation}
which relates the fields at $z$ with their values at the origin of
the $n-$th unit cell. From these two equations one obtains the following
expressions for $h_{a}(z)$ and $f_{a}(z)$: 
\begin{align}
h_{a}(z) & =h_{a}(nd)\frac{\sin(k_{a}(a+nd-z))}{\sin(k_{a}a)}\nonumber \\
 & -h_{a}(nd+a)\frac{\sin(k_{a}(dn-z))}{\sin(k_{a}a)},
\end{align}
and

\begin{align}
f_{a}(z) & =-h_{a}(nd)\frac{\cos(k_{a}(a+nd-z))}{\sqrt{\epsilon_{a}}\sin(k_{a}a)}\nonumber \\
 & +h_{a}(nd+a)\frac{\cos(k_{a}(nd-z))}{\sqrt{\epsilon_{a}}\sin(k_{a}a)}.
\end{align}
Following an analogous procedure for the other dielectric slab, we
write: 
\begin{equation}
\left(\begin{matrix}h_{b}(d(n+1))\\
f_{b}(d(n+1))
\end{matrix}\right)=T_{b}(b)\left(\begin{matrix}h_{b}(nd+a)\\
f_{b}(nd+a)
\end{matrix}\right),
\end{equation}
\begin{equation}
\left(\begin{matrix}h_{b}(z)\\
f_{b}(z)
\end{matrix}\right)=T_{b}(z-nd-a)\left(\begin{matrix}h_{b}(nd+a)\\
f_{b}(nd+a)
\end{matrix}\right).
\end{equation}
Once again, this allows us to obtain expressions for $h_{b}(z)$ and
$f_{b}(z)$ from their values at the interfaces:

\begin{align}
h_{b}(z) & =h_{b}(nd+a)\frac{\sin((d(n+1)-z)k_{b})}{\sin(k_{b}b)}\nonumber \\
 & -h_{b}(d(n+1))\frac{\sin(k_{b}(nd+a-z))}{\sin(k_{b}b)},
\end{align}
and

\begin{align}
f_{b}(z) & =-h_{b}(nd+a)\frac{\cos(k_{b}(d(n+1)-z))}{\sqrt{\epsilon_{b}}\sin(k_{b}b)}\nonumber \\
 & +h_{b}(d(n+1))\frac{\cos(k_{b}(nd+a-z))}{\sqrt{\epsilon_{b}}\sin(k_{b}b)}.
\end{align}

We now impose the continuity of the electric and magnetic fields at
$z=nd+a$ and at $z=nd$, from where we obtain: 
\begin{align}
 & \frac{h(nd)}{\sqrt{\epsilon_{a}}\sin(k_{a}a)}+\frac{h(d(n+1))}{\sqrt{\epsilon_{b}}\sin(k_{b}b)}=\nonumber \\
 & =h(nd+a)\left(\frac{\cos(k_{a}a)}{\sqrt{\epsilon_{a}}\sin(k_{a}a)}+\frac{\cos(k_{b}b)}{\sqrt{\epsilon_{b}}\sin(k_{b}b)}\right),\label{eq: Continuity 1}
\end{align}
\begin{align}
 & \frac{h((n-1)d+a)}{\sqrt{\epsilon_{b}}\sin(k_{b}b)}+\frac{h(nd+a)}{\sqrt{\epsilon_{a}}\sin(k_{a}a)}=\nonumber \\
 & =h(dn)\left(\frac{\cos(k_{b}b)}{\sqrt{\epsilon_{b}}\sin(k_{b}b)}+\frac{\cos(k_{a}a)}{\sqrt{\epsilon_{a}}\sin(k_{a}a)}\right).\label{eq:Continity 2}
\end{align}
Note that the second equation involves both the $(n-1)$ and the $n-$th
unit cells. To simplify these expressions we introduce the following
quantities: 
\begin{align}
A & =\frac{1}{\sqrt{\epsilon_{a}}\sin(k_{a}a)},\label{eq:A}\\
B & =\frac{1}{\sqrt{\epsilon_{b}}\sin(k_{b}b)},\\
C & =A\cos(k_{a}a)+B\cos(k_{b}b),\\
\psi_{n} & =h(nd+a),\\
\phi_{n} & =h(nd).\label{eq:phi_n}
\end{align}
Using this new notation, Eqs. (\ref{eq: Continuity 1}) and (\ref{eq:Continity 2})
become: 
\begin{align}
-A\phi_{n}+C\psi_{n}-B\phi_{n+1} & =0,\\
-A\psi_{n}+C\phi_{n}-B\psi_{n-1} & =0.
\end{align}
We now note that these equations are those defining the SSH model
in real space, i.e. a tight-binding approximation but with energy
dependent hopping parameters $A$ and $B$, and onsite energy $C$.

%

\end{document}